\documentclass{article}
\usepackage[doublespacing]{setspace}

\input{tcilatex}

\begin{document}

\begin{center}
{\LARGE Complementarity and Scientific Rationality}

\bigskip

{\LARGE Simon Saunders}

\medskip

Faculty of Philosophy, University of Oxford

10 Merton St., Oxford OX1 4JJ

\bigskip

\bigskip
\end{center}

\begin{quote}
\textbf{Abstract}: Bohr's interpretation of quantum mechanics has been
criticized as incoherent and opportunistic, and based on doubtful
philosophical premises. If so Bohr's influence, in the pre-war period of
1927-1939, is the harder to explain, and the acceptance of his approach to
quantum mechanics over de Broglie's had no reasonable foundation. But Bohr's
interpretation changed little from the time of its first appearance, and
stood independent of any philosophical presuppositions. The principle of
complementarity is itself best read as a conjecture of unusually wide scope,
on the nature and future course of explanations in the sciences (and not
only the physical sciences). If it must be judged a failure today, it is not
because of any internal inconsistency.

\medskip
\end{quote}

\noindent \textit{Despite the expenditure of much effort, I have been unable
to obtain a clear understanding of Bohr's principle of complementarity}\
(Einstein).

\smallskip \smallskip

\noindent \textit{Bohr's point - and the central point of quantum mechanics
- is that no elementary phenomenon is a phenomenon until it is a registered
(observed) phenomenon }(Wheeler).

\section{A QUESTION POSED}

\noindent It was only in the last quarter of the twentieth century that
alternatives to quantum mechanics finally came of age: only then was it
agreed, by majority (if not quite unanimous) consent, that de Broglie's
pilot-wave theory was indeed empirically equivalent to the non-relativistic
theory; and by the late 1980s that state reduction too might be employed at
ground level in a precise and universal formalism, empirically equivalent to
the standard one.

Recognition of these facts was hard-won, and, in the case of the
state-reduction theory, built on hard labour. But it was otherwise with
pilot-wave theory. The essential ideas for this were laid down by de Broglie
in the 1920s. When Bohm independently discovered the formalism in 1952 he
added only new applications (in particular to the measurement process
itself). Why was it so neglected?

This question becomes more troubling when it is recognized that the view
that did prevail in the inter-war years, Bohr's interpretation of quantum
theory, was flatly in contradiction to the pilot-wave approach.
Embarrassment turns to scandal when it is added that Bohr's view was wedded
to idealism: to the view that there\textit{\ is} no quantum reality, that it
is \textit{the observer} who brings about the result of the measurement. It
must then be explained how the physics community, to an extraordinary degree
of uniformity, could have bought into such an extreme and unnecessary
philosophical position, in support of a theory to which the pilot-wave
theory was a clear counter-example.

The view that Bohr's philosophy \textit{was} so committed - if not to a
Berkelerian idealism, then to anti-realism or to neo-Kantianism - is now
very prevalent (see e.g. \cite{Faye}\cite{Murdoch}). There is plenty of
evidence that Bohr was drawn to neo-Kantianism; there is no doubt that he
saw complementarity as a theory of importance to philosophy as well as
physics. Given which, the view that Bohr's success can \ only be explained,
if explained at all, by \textit{sociological }considerations, of fortuitous
coincidences of philosophical movements, of institutional and collaborative
influences, becomes extremely persuasive: Bohr did stand at the centre of
most of the major discoveries leading to the matrix mechanics; he was
revered by the younger generation of physicists, including Kramers, Pauli,
and Heisenberg (and even Dirac). Cushing was the first to press the question
I am concerned with and was led to just this conclusion \cite{Cushing};
others have since followed him \cite{Beller}.

Or something has gone wrong with our standard of rationality. One can
confront the challenge head-on. One can simply deny that it was reasonable
to buy into pilot-wave theory, at least in the inter-war years. After all it
faces serious difficulties when it comes to relativity,\footnote{%
I have argued elsewhere that the only feasible option here is the Dirac hole
theory (although whether this is really credible, from a realist point of
view, is another matter). Working with field-configurations instead, as
recommended by Valentini \cite{Valentini}, is of no use in making contact
with phenomenology, and worse, if these are the beables, provides no
guarantee that objects will be localized at the macroscopic level, where
they need to be \cite{Saunders}.} and not all is at is seems when it comes
to its account of the observable phenomena (I\ shall come back to this in
due course). Just as important, almost no-one at the time (not even de
Broglie) was prepared to believe that the pilot-wave - a complex-valued
function on configuration space, so a space of enormously high dimension -
could be taken as physically real. And it gets much worse when one takes on
board the implications of extending this idea to the macroscopic level, as
one must, if pilot-wave theory is to get rid of state-reduction. De Broglie
was never prepared to take this step. Although he was perfectly clear that
one could drop non-overlapping \textquotedblleft empty\textquotedblright\
waves from the guidance equation - so providing for an \textquotedblleft
effective\textquotedblright\ state-reduction - he could not countenance the
same treatment when it came to the measurement apparatus and the macroscopic
world (not even \textit{after} reading Bohm's papers \cite{de Broglie2}\cite[%
p.178]{de Broglie}). The pilot-wave in that case is nothing less than a
wave-function to the universe. It was not Bohm in 1952 who was the first to
explore the meaning of that concept, but Everett, in 1957. Who says it was
rational to buy into this picture of reality?\footnote{%
The argument that pilot-wave theory incorporates the same ontology as
Everett's has been made by Brown and Wallace \cite{Brown2}. I would only add
that if it is to be resisted, it is at the price of denying the
intelligibility of the state-reduction approach altogether (so this too
faces a \textquotedblleft problem of rationality\textquotedblright ,
according to the pilot-wave theory).}

Hindsight does not always yield clear vision. Nevertheless, believable or
not, pilot-wave theory is clearly a causal, spacetime theory, of the sort
that Bohr denied was possible. There is undoubtedly a severe tension, if not
outright contradiction, between Bohr's account of quantum mechanics and de
Broglie's. \ And embarrassment at the failure to expose it still turns to
scandal if, as is alleged, Bohr's philosophy was itself incoherent or
philosophically extreme. It is this more limited question that I am
concerned with in the sequel.

\section{HOW TO DEFEND COMPLEMENTARITY}

There have been plenty of attempts in recent years to make sense of Bohr's
writings as a coherent and systematic interpretation of quantum mechanics,
but no account of this sort that I know of is directed to the choices that
were made by the wider community of physicists in the '20s and '30s.%
\footnote{%
Howard's \cite{Howard} is a possible exception, but as he himself says, his
account is a reconstruction rather than an exegesis of Bohr's interpretation.%
} That limits the options. From that perspective, if it is to be defended at
all, Bohr's principle of complementarity, that stands at the heart of his
interpretation, has to make sense as a principle that was plausible and
accessible to his readers at the time. It will not do to read it as based on
esoteric philosophical principles; it will not do to justify it by
mathematical or physical arguments that were then unfamiliar. The answer had
better lie more on the surface of his writings.

What is wrong with complementarity as based on philosophical principles? The
answer is not only that it would then have hardly had much appeal among
physicists; it is that it would have had too broad a scope, and pilot-wave
theory refutes it. As we shall see, the principle needs to be read as
anchored in the phenomena if it is to stand any hope of escaping this
objection. Only on one point was it reasonable to commit to any \textit{a
priori} (or dogmatic) principle, in the absence (at the time) of any worked
out view to the contrary, and that was the irreplaceability of classical
concepts (agreed on by Bohr, Einstein and de Broglie). The old quantum
theory was of course also based on classical concepts; Bohr was as much
concerned to make sense of the old quantum theory as the new.

\textit{Broadly} philosophical presuppositions can be allowed, so long as
they were genuinely community-wide. Anti-realism and idealism are not in
this category, but a mildly operationalist outlook is.\footnote{%
The notion of \textquotedblleft the observer\textquotedblright\ as a central
concept of special relativity was also common ground to Bohr and his
readers. Bohr often drew attention to it in his discusions of quantum
mechanics. But the parallel was not essential to Bohr's arguments (it was
the notion of a frame of rerence that played the greater role), and I shall
not consider it here.} Bohr was concerned with experiments and with the
operational definitions of concepts, as were most of his contemporaries. A
broad and shallow form of operationalism is perfectly compatible with
realism (\textquotedblleft realism with a cautious face\textquotedblright ).

Finally, since our concern is with Bohr's influence on the wider scientific
community in the inter-war period, it is only his views and writings at this
time that matter. Three stand out as especially important. The first is his
address to the International Congress of Physics in Como, Italy, in 1927
(the \textit{Como lecture}), published in \textit{Nature\ }the following
year as \textquotedblleft The quantum postulate and the recent development
of atomic theory\textquotedblright\ (and, in German, in \textit{Die
Naturwisssenschaften}). It was the centre piece of his collection \textit{%
Atomtheorie und Naturbeschreibung }published in 1931, translated as \textit{%
Atomic Theory and Human Knowledge} in 1934 \cite{Bohr1}. In the Como lecture
Bohr first set out his theory of complementarity, following two years of
almost complete public silence (the two years that covered the explosive
discovery of the new mechanics). The second is the preface to this
collection, the \textquotedblleft Introductory Survey\textquotedblright\ 
\cite[pp.1-24]{Bohr1}, first published in 1929 (in Danish)\ as an
accompaniment to the Danish translations of these articles. In the German
and English translations this was read almost as widely as the Como lecture
itself. And the third is Bohr's response to Einstein, Podolski, and Rosen
(the EPR argument), published in 1935 \cite{Bohr2}; that followed several
years of debate with Einstein over foundations and in effect marked their
conclusion. Any account of Bohr's interpretation of quantum mechanics that
is not clearly embodied in these three texts is worthless for our purposes.

Better still is an account consistent with these and with Bohr's own
evaluation of his thinking in this period. That we have in his contribution
to the Schilpp volume in the \textit{Library of Living Philosophers }devoted
to Einstein, published in 1949 \cite{Bohr3}; there Bohr reappraised both the
Como lecture and his discussions with Einstein. In this spirit one other
article is worth special mention, \textquotedblleft Quantum physics and
philosophy - causality and complementarity\textquotedblright\ published\ in
1958 \cite{Bohr4}. It was Bohr's last presentation of complementarity.
According to his son and executor Aage Bohr, he felt that there he expressed
some of its key concepts more clearly and concisely than he had elsewhere.
Bohr died in 1962.

My claim is that in these writings Bohr is most clearly and consistently
read as a realist, albeit of an operational persuasion; that his goal was to
present a framework in which quantum phenomena were to be analyzed in
classical terms; and that he argued for this framework in terms of
principles, and specifically the principle of complementarity, as \textit{%
scientific} principles, broadly empirical in scope, rather than
philosophical ones, that stood independent of idealism or neo-Kantianism (or
any other school in philosophy). And in these respects I maintain he was
largely successful.

None \ of this is to say that it was reasonable, \textit{circa }the late
1920s and '30s, to embrace Bohr's views and to reject de Broglie's and
Einstein's; but it is a step in the right direction. I shall have a further
comment to make on this at the end.

\section{THE COMO LECTURE}

In 1927 Bohr's point of departure was the \textit{quantum postulate}. A
principle of that name had long been familiar to the old quantum theory, as
defined by the Bohr-Sommerfeld quantization rules (the principle that any
change in action, with the units of angular momentum, must be an integral
multiple of Planck's constant $h)$. Of quantum mechanics, he began:

\begin{quotation}
\noindent \noindent Its essence may be expressed in the so-called quantum
postulate, which attributes to any atomic process an essential
discontinuity, or rather individuality, completely foreign to the classical
theories and symbolized by Planck's constant of action. \cite[p.53]{Bohr1}
\end{quotation}

\noindent Bohr immediately went on to say that the quantum postulate
\textquotedblleft implies a renunciation as regards the causal spacetime
coordination of atomic processes\textquotedblright .

Both claims were by then contentious, given that Schr\"{o}dinger had deduced
the quantization rule as a consequence of boundary conditions for the
solutions of a continuous wave equation only the previous year. It may be
that Bohr was convinced, with the experience of the failure of the Bohr,
Kramers and Slater theory just behind him, that if energy and momentum were
conserved in individual processes then quantum jumps were just as
unavoidable in wave mechanics as in matrix mechanics (Schr\"{o}dinger's wave
function and Slater's virtual radiation field were closely allied). Later on
in the Como lecture Bohr spoke of wave mechanics as \textquotedblleft a
symbolic transcription\textquotedblright\ that is \textquotedblleft only to
be interpreted by an explicit use of the quantum postulate\textquotedblright
. On two occasions he spoke of the postulate as \textquotedblleft
irrational\textquotedblright\ (as in \textquotedblleft ...we meet...the
inevitability of the feature of irrationality characterizing the quantum
postulate\textquotedblright ). This makes the abandonment of causal
spacetime descriptions look like an assumption from the very beginning.

But the argument that followed was more circumspect. There was a special
sense in which the causal, spacetime idea of explanation was to be weakened.
Here is the argument on its first appearance:

\begin{quotation}
\noindent Now, the quantum postulate implies that any observation of atomic
phenomena will involve an interaction with the agency of observation not to
be neglected. Accordingly, an independent reality in the ordinary physical
sense can neither be ascribed to the phenomena nor to the agencies of
observation.
\end{quotation}

\noindent Bohr argues from the quantum postulate, understood as implying an
ineliminable interaction on observation, to a \textquotedblleft
no-separation\textquotedblright\ principle - that the object of observation
is inseparably bound up with the experimental context. Similar claims may be
found in all his subsequent writings on quantum mechanics.

Bohr continued:

\begin{quotation}
\noindent After all, the concept of observation is in so far arbitrary as it
depends upon which objects are included in the subsystem to be observed.
Ultimately, every observation can, of course, be reduced to our sense
perceptions. The circumstance, however, that in interpreting observations
use has always to be made of theoretical notions entails that for every
particular case it is a question of convenience at which point the concept
of observation involving the quantum postulate with its inherent
"irrationality" is brought in.
\end{quotation}

\noindent Bohr here and subsequently is certainly preoccupied with
\textquotedblleft observation\textquotedblright , with experiments - on what
can be said of the microscopic realm on the basis of experiments. To this
extent his philosophy was broadly operationalist. On the other hand his
position is far from positivist, in Mach's sense: Bohr is clear that
observations, whether or not they are reducible to sense impressions (as it
happens he grants that they are), must be expressed in terms of concepts -
they are \textit{interpreted} - and in this precisely where one puts the
boundary between the observed system and the context of the observation is
somewhat arbitrary. Bohr repeatedly spoke of \textquotedblleft
measurement\textquotedblright\ (\textquotedblleft agencies of
measurement\textquotedblright ) in the sequel: the boundary at issue for
Bohr as much marks the distinction between the context of the experiment and
the object under investigation, as between \textquotedblleft the
observer\textquotedblright\ and \textquotedblleft the
observed\textquotedblright . In his later writings it was the former that
was increasingly to the fore.\footnote{%
For a typical example: \textquotedblleft We must recognize that a
measurement can mean nothing else than the unambiguous comparison of some
property of the object under investigation with a corresponding property of
another system, serving as a measuring instrument, and for which this
property is directly determinable according to its definition in everyday
language or in the terminology of classical physics.\textquotedblright \cite[%
p.19]{Bohr5}}

He immediately continues:

\begin{quotation}
This \ situation has far-reaching consequences. On one hand, the definition
of the state of a physical system, as ordinarily understood, claims the
elimination of all external disturbances. But in that case, according to the
quantum postulate, any observation will be impossible, and, above all, the
concepts of space and time lose their immediate sense. On the other hand, if
in order to make observation possible we permit certain interactions with
suitable agencies of measurement, not belonging to the system, an
unambiguous definition of the state of the system is no longer possible, and
there can be no question of causality in the ordinary sense of the word. The
very nature of the quantum theory thus forces us to regard the space-time
co-ordination and the claim of causality, the union of which characterizes
the classical theories, as \textit{complementary} but exclusive features of
the description, symbolizing the idealization of observation and definition
respectively. \cite[p.54, emphasis original]{Bohr1}
\end{quotation}

\noindent This is the first time that the word \textquotedblleft
complementary\textquotedblright\ was used by Bohr.

The simplest reading of this passage is more strongly operationalist: the
concepts of space and time have no meaning independent of a context of
observation. As a philosophical doctrine it will therefore apply equally to
classical physics. The difference, for microscopic quantum phenomena, is
that the act of observation cannot under any circumstances be neglected.
Because of the quantum postulate, there is an irreducible coupling between
apparatus and measured system, that cannot be made arbitrarily small.%
\footnote{%
Understood as a statement about energy or momentum transfers, this point is
not entirely general. What is entirely general is that, for a given basis,\
the presence or absence of \textit{entanglement} can never be neglected.
Much hangs on this distinction.} A microscopic system to which spacetime
coordinates can be assigned can therefore never be considered a closed
system, not even as an idealization. So equations of motion in the customary
form are not available; what equations may be found for it will not conserve
energy and momentum; no \textquotedblleft causal\textquotedblright\
description is possible. In this sense spacetime coordination and causality
cannot be combined. Further, at least in a number of important examples, the
reciprocal nature of this limitation can be quantified by means of the
uncertainty relations.

This reading is consistent with the rest of the Como lecture. There and in
his later writings Bohr repeatedly gave examples to show that the attempt to
give an operational meaning to the spatiotemporal coordinates of a
phenomenon leads to an uncontrollable flow of momentum and energy into and
out of the system, of just such an amount as to satisfy the uncertainty
relations. As of the Como lecture the foundation of the latter was the de
Broglie relations: in them the \textquotedblleft fundamental contrast
between the quantum of action and the classical concepts is immediately
apparent\textquotedblright . Momentum and energy (the basis of a
\textquotedblleft causal\textquotedblright\ description) is thereby related
to wavelength and frequency; with that the uncertainty relations (or near
neighbours of them) follow immediately:

\begin{quotation}
\noindent At the same time, the possibility of identifying the velocity of
the particle with the group-velocity indicates the field of application of
space-time pictures in the quantum theory. Here the complementary character
of the description appears, since the use of wave-groups is necessarily
accompanied by a lack of sharpness in the definition of period and
wave-length, and hence also in the definition of the corresponding energy
and momentum as given by [the de Broglie relations]. \cite[p.59]{Bohr1}
\end{quotation}

\noindent Bohr's derivation of the uncertainty relations depended only on
the de Broglie relations and the concepts of elementary (linear) wave theory.

This point is worth emphasizing, for whilst Bohr, as of the time of the Como
lecture, had finally taken on board what he had always regarded as the most
profound paradox of the quantum theory - the wave-particle duality - he had
yet to absorb even quite superficial features of the new mechanics. As he
apologetically prefaced his address:

\begin{quotation}
\noindent \noindent \lbrack A]mong the audience there will be several who,
due to their participation in the remarkable recent development, will surely
be more conversant with details of the highly developed formalism than I am.
Still, I shall try, by making use only of simple considerations and without
going into any details of the technical mathematical character, to describe
to you a certain general point of view \textit{which I believe is suited to
give an impression of the general trend of the development of the theory
from its very beginning} and which I hope will be helpful in order to
harmonize the apparently conflicting views taken by different scientists.\ 
\cite[p.52, emphasis mine]{Bohr1}
\end{quotation}

\noindent It is most unlikely that Bohr meant by the \textquotedblleft very
beginning\textquotedblright\ the beginning of the \textit{new} quantum
mechanics (no more than two years old). Evidently, he was as much concerned,
in the Como lecture, with the trend of development of the \textit{old }%
quantum theory, as with the new.

Our reading of Bohr to this point is relatively uncontroversial, but the
initial steps of Bohr's argument remain obscure. What, precisely, is the
\textquotedblleft quantum postulate\textquotedblright , and why does it lead
to what I\ am calling the \textquotedblleft no-separation\textquotedblright\
principle - the doctrine that neither the agency of observation nor the
object observed can be ascribed \textquotedblleft independent reality in the
ordinary physical sense\textquotedblright ? Bohr presented this conclusion
as an immediate consequence of the impossibility of neglecting the
measurement interaction, but it is not clear if this is a reference to the
\textquotedblleft individuality\textquotedblright\ of an atomic process -
whatever, precisely, that may mean - or its \textquotedblleft essential
discontinuity\textquotedblright . But either way, Bohr is evidently taking
it as an \textquotedblleft external disturbance\textquotedblright\ to a
system (for only then does it imply that on measurement a system \textit{%
cannot }be free of any external disturbance). It is tempting to go on to
take Bohr to mean \textquotedblleft disturbance\textquotedblright\ in its
normal sense, as a causal physical process. With that one is led very
quickly to simple-minded \textit{disturbance\ theory of measurement}. The
observed system is \textit{disturbed} by its interaction with the measuring
apparatus, so it can no longer be treated as isolated (and the
energy-momentum conservation laws will no longer hold for it, so
\textquotedblleft causality\textquotedblright\ is violated).

The disturbance theory of measurement had the virtues of simplicity and
clarity, and it was certainly popular; it figured in many of the early texts
in quantum mechanics; but it falls foul to obvious objections. Why not seek
to correct for the disturbance, as one does classically, in cases where the
measurement interaction is not in fact negligible? (obviating the
no-separation principle). Why not, to this end, include the measuring
apparatus in with the measured system, and model the interaction between the
two directly, in quantum mechanical terms? We know of course of one answer
to this question: that if we apply the unitary equations of motion to the
apparatus as well as to the system measured, we are led to a description of
no \textit{one }definite outcome at all - apparently, to nonsense. But Bohr
did not begin with the abstract formalism; he did not acknowledge the
problem of measurement as such. His point was not: here is the problem of
measurement, to solve it we have to insist on X. Bohr's point is: here is
the quantum postulate and some philosophical or physical principles; deduce
X (and from X, one might hope, solve or dissolve the problem of
measurement). So what, according to Bohr, is wrong with applying the
ordinary equations (suitable for a closed system) to the apparatus and
observed system taken together? Bohr can continue to insist that that
description, if cast in terms of space and time coordinates, must itself be
observed by some agency (to \ give \textquotedblleft
sense\textquotedblright\ to the coordinations), but why an \textit{outside}
agency? Cannot the universe be observed from within? Is he bound by some
neo-Kantian injunction or what-not against global descriptions?

On this reading Bohr is on awkward ground. His thesis is in danger of
becoming overtly philosophical, and hostage to philosophical arguments that
may take unforeseen directions. It is hardly what he intends. He is trying
to do justice \textquotedblleft to the general trend of the development of
the theory\textquotedblright\ (including the old quantum theory) as
recognized by physicists, not in accordance with arcane metaphysical
principles. To insist on clear operational meanings to physical concepts is
one thing; on Kantian bounds of sense is quite another.

All of this flows from the simple-minded picture of the experiment as
introducing a disturbance in the object measured. If this were the whole
story, Bohr, on review of Heisenberg's operational analysis of the
uncertainty relations in terms of a disturbance on measurement, would not
have continued as he did:

\begin{quotation}
The essence of this consideration is the inevitability of the quantum
postulate in the estimation of the possibilities of measurement. \textit{A
closer investigation of the possibilities of definition would still seem
necessary in order to bring out the general complementary character of the
description}. Indeed, a discontinuous change of energy and momentum during
observation could not prevent us from ascribing accurate values to the
space-time coordinates, as well as to the momentum-energy components before
and after the process. The reciprocal uncertainty which always affects the
values of these quantities is, as will be clear from the proceeding
analysis, essentially an outcome of the limited accuracy with which changes
in energy and momentum can be \textit{defined}, when the wave-fields used
for the determination of the space-time coordinates of the particle are
sufficiently small. \cite[p.63, emphasis mine]{Bohr1}
\end{quotation}

\noindent Bohr went on to speak of \textquotedblleft the complementarity of
the possibilities of \textit{definition}\textquotedblright , emphasizing
that \textquotedblleft the agreement between the possibilities of \textit{%
observation} and those of \textit{definition} can be directly
shown\textquotedblright\ (emphasis mine); that sets clear limits to any
positivist elements in Bohr's operationalism. But the most decisive reason
to reject the assimilation of complementarity as of this point to a
disturbance theory of measurement is given, not by Bohr, but by Heisenberg,
in a note added in proof to his paper on the uncertainty relations:

\begin{quotation}
\noindent Bohr has brought to my attention that I have overlooked essential
points in the course of several discussions in this paper. Above all, the
uncertainty in our observation does not arise exclusively from the
occurrence of discontinuities, but is tied directly to the demand that we
ascribe equal validity to the quite different experiments which show up in
the corpuscular theory on one hand, and in the wave theory on the other. 
\cite[p.198]{Heisenberg1}
\end{quotation}

\noindent That is precisely how Bohr went on to illustrate complementarity
at the beginning of the Como lecture, first in terms of the wave-particle
duality for light, and then for matter:\ 

\begin{quotation}
\noindent Just as in the case of light, we have consequently in the question
of the nature of matter, so far as we adhere to classical concepts, to face
an inevitable dilemma which has to be regarded as the very expression of
experimental evidence. In fact, here again we are not dealing with
contradictory but with complementary pictures of the phenomena, which only
together offer a natural generalization of the classical mode of
description. \cite[p.56]{Bohr1}
\end{quotation}

Complementarity, on its first appearance, was thus a thesis concerning the
contextuality of the phenomenon to the experiment, as expressed by classical
concepts, under a reciprocal latitude in definition as follows from the de
Broglie relations; and of the agreement between this and a corresponding
reciprocity in their simultaneous measurability. He never changed his views
on these matters.

But not all of Bohr's assumptions were properly in evidence in the Como
lecture. Bohr spoke of adhering to the classical concepts - but why should
we? The challenge was made shortly after by Schr\"{o}dinger in
correspondence: that interesting as the limitations of the classical
concepts were, as subject to the uncertainty relations

\begin{quotation}
\noindent \noindent \lbrack I]t seems to me imperative to demand the
introduction of new concepts, in which this limitation no longer occurs.
Since what is unobservable in principle should not at all be contained in
our conceptual scheme, it should not be representable in terms of the
latter. In the adequate conceptual scheme it ought no more to seem that our
possibilities of experience are restricted through unfavourable
circumstances \cite[p.465]{Bohr6}
\end{quotation}

\noindent Bohr could not have been more cool in his reply:

\begin{quotation}
\noindent \noindent I am scarcely in complete agreement with your stress on
the necessity of developing \textquotedblleft new\textquotedblright\
concepts. Not only, as far as I can see, have we up to now no clues for such
a re-arrangement, but the \textquotedblleft old\textquotedblright\
experiential concepts seem to me to be inseparably connected with the
foundation of man's power of visualising \cite[p.465]{Bohr6}
\end{quotation}

\noindent He is on shaky ground, however. It was not so long before that
Euclidean geometry was supposed to be the only visualizable geometry, the
existence of mathematical schemes for non-Euclidean geometries
notwithstanding. Bohr's position, at this point, is dogmatic.

If a point of dogma, better state it at the beginning of any argument for
complementarity, and better free it from any reliance on dubious empirical
claims about our \textquotedblleft powers of visualization\textquotedblright
. It came in the very first paragraph of the \textquotedblleft Introductory
Survey\textquotedblright\ to his \textit{Atomic Theory and the Description
of Nature}, written in 1929:

\begin{quotation}
\noindent Only by experience itself do we come to recognize those laws which
grant us a comprehensive view of the diversity of phenomena. \noindent As
our knowledge becomes wider, we must always be prepared, therefore, to
expect alterations in the points of view best suited for the ordering of our
experience. In this connection we must remember, above all, that, as a
matter of course, all new experience makes its appearance within the frame
of our customary points of view and forms of perception. \cite[p.1]{Bohr1}
\end{quotation}

\noindent Later on in the same survey the point was made again, this time
with reference to \textquotedblleft our customary ideas or their direct
verbal expressions\textquotedblright . This commitment to classical concepts
on the basis of their role in ordinary language figured repeatedly in Bohr's
subsequent writings. It was, for Bohr, the rational for a far more sweeping
commitment. He immediately continued:

\begin{quotation}
\noindent No more is it likely that the fundamental concepts of the
classical theories will ever become superfluous for the description of
physical experience. The recognition of the indivisibility of the quantum of
action, and the determination of its magnitude, not only depend on an
analysis of measurements based on classical concepts, but it continues to be
the application of these concepts alone that makes it possible to relate the
symbolism of the quantum theory to the data of experience. \cite[p.16]{Bohr1}
\end{quotation}

\noindent The jump from the necessity and unrevisability of the concepts of 
\textit{everyday experience} to that of the fundamental concepts of \textit{%
classical theories} was unsubstantiated, however.\footnote{%
One might wonder if, far from exhausting the concepts available in the
description of our experience, the use of classical concepts is even so much
as consistent with quantum mechanics. This question, of whether classical
concepts, taken individually, could so much as be employed in the quantum
domain, had guided Heisenberg in his discovery of the uncertainty relations;
his conclusion was that they could (and that only their \textit{simultaneous}
deployment was circumscribed) so Bohr's piecemeal use of them did at least
have Heisenberg's sanction. In Heisenberg's words: \textquotedblleft \textit{%
All concepts which can be used in classical theory for the description of a
mechanical system can also be defined exactly for atomic processes in
analogy to the classical concepts.}\textquotedblright\ \cite[p.68, emphasis
original]{Heisenberg1}.\ (In point of fact, the claim at this level of
generality runs very quickly into trouble. Shortly after, Jordan and Dirac
both noted the difficulties of giving any meaning to the time or phase as
self-adjoint operators obeying canonical commutation relations with \ the
energy, if the latter is to have a point spectrum. Mathematically, the
difficulty is that no quantization procedure has been found in which the
full symmetry group of classical phase space, the symplectic group, can be
implemented as a group of unitary transformations on a Hilbert space.)} But
almost no-one apart from Schr\"{o}dinger saw this as a weakness of Bohr's
interpretation; and certainly not Einstein, his principal critic.

\section{BOHR'S RESPONSE TO THE EPR ARGUMENT}

The early 1930s were the must crucial years for the interpretation of
quantum mechanics. They followed much publicized and visible encounters
between Einstein and Bohr on the subject of foundations (in particular at
the 6th Solvay Conference of 1930). With the appearance of the first
comprehensive introductions to the subject, by Pauli, Kramers, Jordan and
Dirac, and with time to assimilate the new formalism, a deeper appreciation
of its paradoxes was in the air. Things came to a head in 1935: in that year
a number of criticisms of the new mechanics were published, by von Laue, Schr%
\"{o}dinger, and (with Podolski and Rosen) Einstein. It was Einstein's
parting shot: he left Germany for America in 1933, never to return.

The view is very widely held that Bohr, without admitting it, shifted his
position markedly in the face of these developments, and above all the EPR
argument. That would be a damaging admission, if true; for not only would
Bohr stand revealed as an opportunist, it would show that the community
embraced quite distinct orthodoxies without even realizing it. Fortunately,
however, whilst there undoubtedly were shifts in Bohr's position, they
effected his argumentative strategy more than its substance. On substance
the changes were subtle.

This claim needs to be justified. Bohr frequently remarked on the value he
placed on his discussions with Einstein, almost all of which took place
before Einstein's departure for America. Here is a lesson he said he learned
early from them:

\begin{quotation}
\noindent The extent to which renunciation of the visualization of atomic
phenomena is imposed upon us by the impossibility of their subdivision is
strikingly illustrated by the following example to which Einstein very early
called attention and often has reverted. If a semi-reflecting mirror is
placed in the way of a photon, having two possibilities for its direction of
propagation, the photon may either be recorded on one, and only one, of two
photographic plates situated at great distances in the two directions in
question, or else we may, by replacing the plates by mirrors, observe
effects exhibiting an interference between the two reflected wave-trains. In
any attempt of a pictorial representation of the behaviour of the photon we
would, thus, meet with the difficulty: to be obliged to say, on the one
hand, that the photon always chooses one of the two ways and, on the other
hand, that it behaves as if it had passed both ways. \cite[p.221]{Bohr3}
\end{quotation}

\noindent It is an early example of a \textit{delayed-choice} experiment.
One must change the description of a system in the past, needed to explain a
measurement, depending on which of two measurements one chooses to make
later on.

In one respect this is worse than any non-locality in space, as was shortly
to be demonstrated by the EPR argument \cite{EPR}: it is an action of the
present on the past.\footnote{%
However it was widely accepted that retrodictions have a rather different
status from predictions in quantum mechanics, in view of the fact that - say
from successive measurements of position of arbitrary accuracy - one can
defeat the uncertainty relations. According to Bohr in the Como lecture, in
such cases we deal with an \textquotedblleft abstraction, from which no
unambiguous information can be obtained\textquotedblright .} It is a case in
which the phenomenon is contextualized to the experimental conditions; it
illustrates Bohr's \textquotedblleft no-separation\textquotedblright\
principle. Bohr had, moreover, already met with attempts by Einstein to
extend this to non-locality in a predictive sense (I shall come back to
these in a moment). If the paper of Einstein\textit{\ et al} really came as
an \textquotedblleft onslaught...as a bolt from the blue\textquotedblright ,
as Rosenfeld later said \cite[p.142]{Rosenfeld}, the ideas were by no means
entirely unfamiliar to Bohr (which does not of course mean that he had
anticipated the argument). Nor did it take him long to respond to it, by his
standards - little over a month - in a short note in \textit{Nature} \cite%
{Bohr9}; this, almost verbatim, was the core of the much longer reply he
published in the \textit{Physical Review} near the end of the year \cite%
{Bohr2}. In the latter he began with well-known experiments that he had
already used as examples of complementarity. According to Bohr, the EPR
argument \textquotedblleft does not actually involve any greater intricacies
than the simple examples discussed above\textquotedblright . If Bohr saw
anything new in the EPR\ argument, he did not acknowledge it. Yet for
Einstein it was conclusive proof that quantum mechanics was incomplete, a
view that he held to the end of his life.

The EPR argument, recall, rested on a sufficiency condition for a quantity
to be counted an \textquotedblleft element of reality\textquotedblright .
The condition was that the quantity can be predicted with certainty
\textquotedblleft without in any way disturbing the system\textquotedblright
. Depending on which of two experiments was performed on one system, and on
the outcome obtained, the value of one or other of two non-commuting
quantities associated with a second system could be predicted with
certainty. Since this is so even in the absence of any interaction between
the two systems, the sufficiency condition is satisfied; so \textit{both}
must be elements of reality. But they could not both be represented as such
by any single quantum state: quantum mechanical description is therefore
incomplete.

The argument turns on the key concepts of \textquotedblleft
disturbance\textquotedblright\ and \textquotedblleft
interaction\textquotedblright\ in the very context - what Einstein \textit{%
et al} called \textquotedblleft reduction of the wave
packet\textquotedblright\ - that was so critical to Bohr's interpretation
(bringing in \textquotedblleft the quantum postulate\textquotedblright ).
Naturally, therefore, according to Bohr of the Como lecture, it is a context
in which \textquotedblleft an independent reality in the ordinary physical
sense can neither be ascribed to the phenomena nor to the agencies of
observation.\textquotedblright\ To complete the line of thought he expressed
then, this is because the quantum postulate implies that \textit{there is}
an interaction between agency of observation and object (one that is \textit{%
not to be neglected}). What is needed, then, if Bohr is to be consistent
with his creed, is to bite the bullet and admit there is still some kind of
an interaction even if it is not of the usual sort (and, he might have
added, even if it acts at a distance\footnote{%
There were special difficulties in treating the EPR\ state dynamically (in
contrast to Bohm's later version in terms of corelated spins). Neither in
the EPR\ paper nor in Bohr's reply was any mention made of locality. But the
potential non-local character of the influence Bohr spoke of must have been
obvious, given Einstein's previous criticisms of quantum mechanics.}). That
is just what he said:

\begin{quotation}
\noindent Of course there is in a case like that just considered no question
of a mechanical disturbance or the system under investigation during the
last critical phase of the measuring procedure. But even at this stage there
is essentially the question of \textit{an influence on the very conditions
which define the possible types of predictions regarding the future
behaviour of the system}. Since these conditions constitute an inherent
element of the description of any phenomenon to which the term
\textquotedblleft physical reality\textquotedblright\ can be properly
attached, we see that the argumentation of the mentioned authors does not
justify their conclusion that quantum-mechanical description is essentially
incomplete. \cite[p.700, emphasis original]{Bohr2}
\end{quotation}

\noindent (In his earlier one-page reply, Bohr used the phrase
\textquotedblleft ...no question of a direct mechanical
interaction....\textquotedblright .)

It is a wordier version of Bohr's no-separation principle, but now quite
clearly \textit{divorced} from the disturbance picture of measurement. It is
the principle that any physically real phenomenon must be specified under
definite experimental conditions, so any change in the latter must lead to a
change in the former, even if no ordinary interaction is involved. To put it
in spacetime terms (which neither Bohr nor Einstein \textit{et al} had,
given that the EPR\ state was defined at only a single instant of time), it
is not as though one can hold a part of the phenomenon, the remote part,
constant, whilst varying the experimental conditions of the local part of
the phenomenon - this would be to try to visualize the phenomenon in
accordance with causal spacetime concepts; it would be to ignore the
\textquotedblleft individual\textquotedblright\ nature of a quantum
phenomenon.

At this point the strain of not interpreting Bohr's non-mechanical
interaction as entanglement, and the quantum postulate as state-reduction,
becomes well-nigh intolerable, but still it should be resisted. It is not
only that he never accepted these identifications, it was that for Bohr,
formal concepts like entanglement and state reduction could \textit{never}
have been explanatory (the formalism was \textit{only} an abstract
calculus). To put the quantum postulate in terms of state reduction is to
look at Bohr's theory from the wrong direction.

If we stay with Bohr's own terms, there is not a great deal to add in reply
to Einstein \textit{et al} - unless it is to illustrate how Bohr's
no-separation principle had functioned all along, in the familiar cases he
had already analyzed in his Como lecture. That is precisely what the bulk of
his reply in the \textit{Physical Review} contained. But in the course of it
he did refine his position in certain respects, and he took the opportunity,
naturally enough, to put the matter more as he had done in his Introductory
Survey\ of 1929:

\begin{quotation}
\noindent While, however, in classical physics the distinction between
object and measuring agencies does not entail any difference in the
character of the description of the phenomena concerned, its fundamental
importance in quantum theory, as we have seen, has its root in the
indispensable use of classical concepts in the interpretation of all proper
measurements, even though the classical theories do not suffice in
accounting for the new types of regularities with which we are concerned in
atomic physics. In accordance with this situation there can be no question
of any unambiguous interpretation of the symbols of quantum mechanics other
than that embodied in the well-known rules which allow us to predict the
results to be obtained by a given experimental arrangement described in a
totally classical way. \cite[p.701]{Bohr2}
\end{quotation}

\noindent Here Bohr is crystal-clear. His interpretation is of the phenomena
in terms of classical \textit{concepts}, even though no classical\textit{\
theory} can account for such regularities. This was the heart of what was
really innovative about Bohr's principle of complementarity: how could one
describe regularities classically, when they could not be described by any
classical theory?

Before coming to that, we should take note of what is genuinely new in
Bohr's statement over and above that in the \textquotedblleft Introductory
Survey\textquotedblright : it is his insistence that the experimental
arrangement be described in a \textit{totally} classical way, meaning there
was no reciprocal latitude needed in any of the classical concepts involved
(the uncertainty relations do not apply).\footnote{%
In the Como lecture Bohr did suggest on at least one clear occasion that the
agency of measurement could partake of quantum mechanical uncertainties (%
\cite[p.66]{Bohr1}).} Use of constraints on the possible \textquotedblleft
latitudes\ of definition\textquotedblright\ characterizes rather the
\textquotedblleft quantum mechanical description\textquotedblright . Thus
if, in an experiment to measure the position of a particle (using a rigidly
mounted diaphragm), one wishes instead to control for the momentum of the
diaphragm, then it must, \textquotedblleft as regards its position relative
to the rest of the apparatus, be treated, like the particle traversing the
slit, as an object of investigation, in the sense that the quantum
mechanical uncertainty relations regarding its position and momentum must be
taken explicitly into account.\textquotedblright\ (\cite[p.698]{Bohr2}).

One can read the shift as reflecting the need for the von Neumann
\textquotedblleft cut\textquotedblright , but again this is to put it in
terms quite foreign to Bohr. It is rather a matter of recognizing that
eventually one must make use of an apparatus whose reaction to the process
of measurement cannot itself be controlled. There must always come a point
where it is impossible to keep track of any energy and momentum flows
between the apparatus and the object that is measured (as the change in
momentum of the apparatus, let alone the change in position - relative to
what? - become totally inaccessible). One way of making the point is by
insisting that the uncertainty relations are not to be applied to the
apparatus.\footnote{%
See Di\'{o}si \cite{Diosi}, for a technical treatment along these lines.}
Along the way, insofar as Bohr has to draw a definite classical-quantum
distinction, it is a convenient tidying-up exercise: why not draw it at the
same place as the apparatus-object distinction?\footnote{%
Howard calls this the \textquotedblleft coincidence
interpretation\textquotedblright , and goes on to question it, suggesting,
even, that subatomic particles might have been counted by Bohr as measuring
instruments, so long as they are assigned the right spectrum of classical
properties \cite{Howard}. (My disagreement with him should be clear from the
sequel.)}

But it is not really needed. We can replace it, as an expression of our
pragmatic situation, by the stipulation that the conditions of an experiment
must ultimately involve rigid connections to bodies of arbitrarily large
mass. In that case the uncertainty relations, for the latter bodies, become
irrelevant (so long as there is non-zero lattitude in \textit{both} position
and momentum).\footnote{%
The limit of infinite mass is of course singular. But see Dickson \cite%
{Dickson3}, for a rather different view of the matter.} Bohr admitted as
much when he remarked that the freedom of choice in the divide between
quantum and classical was restricted to \textquotedblleft a region where the
quantum mechanical description of the process concerned is effectively
equivalent with the classical description\textquotedblright\ \cite[p.701]%
{Bohr2}, and later, when he said that the requirements of unambiguous
description of the apparatus \textquotedblleft is secured by the use, as
measuring instruments, or rigid bodies sufficiently heavy to allow a
completely classical account of their relative positions and
velocities\textquotedblright\ \cite[p.3]{Bohr4}.

Another tidying-up operation came shortly after his reply to EPR, and was
more explicitly terminological. Recognizing, as Bohr may not have
appreciated before the EPR argument, that the contextualizing of the
phenomenon embodied in the no-separation principle had to be freed much more
explicitly from any causal concepts, it would be handy to devise a
terminology in terms of which the choice of experimental arrangement
strictly does \textit{not }disturb the phenomenon. But that is quite easy to
do. As he later reported his proposal, made at Warsaw in 1938:

\begin{quotation}
\noindent I warned especially against phrases, often found in the physics
literature, such as \textquotedblleft disturbing of phenomena by
observation\textquotedblright , or \textquotedblleft creating physical
attributes to atomic objects by measurement\textquotedblright . Such
phrases, which may serve to remind us of the apparent paradoxes in quantum
theory, are at the same time apt to cause confusion, since words like
\textquotedblleft phenomena\textquotedblright\ and \textquotedblleft
observations\textquotedblright , just as \textquotedblleft
attributes\textquotedblright\ and \textquotedblleft
measurements\textquotedblright , are used in a way hardly compatible with
common language and practical definition.

As a more appropriate way of expression I advocated the application of the
word \textit{phenomenon} exclusively to refer to the observations obtained
under specified circumstances, including an account of the whole
experimental arrangement. \cite[p.237-38, emphasis original]{Bohr3}
\end{quotation}

\noindent With that talk of any disturbing of the phenomenon by a change in
experimental context becomes \textit{literally} false, in fact a logical
contradiction. If the kind of experiment is changed we have a completely
different phenomenon.

Does it follow that there is no longer a role for the idea that quantum
measurements disturb the system measured? Not in the least: it is essential,
to get the whole doctrine of complementarity off the ground, that there are
indeed mutually exclusive experimental arrangements. We are not talking of
logical incompatibility; the point is not that if one performs a two-slit
experiment one cannot at the same time perform a diffraction-grating
experiment (a contradiction in terms, even though the observables that are
measured commute); the incompatibility rather derives from a physical
principle that implies an experiment to measure one classical quantity
thereby excludes the possibility of simultaneously measuring another. Here
the notion of an irreducible disturbance works perfectly well as limited to
a purely local action.\footnote{%
In fact, by microcausality, this is guaranteed.} As Bohr went on to explain,
in defending the completeness of the quantum mechanical description:

\begin{quotation}
\noindent On the contrary this description,\footnote{%
Here, for once, we should take Bohr as talking about the quantum state (the
EPR state), understanding its decomposition with respect to the position
basis or the momentum basis as illustrating the complementary descriptions
that can be given of the system.} as appears from the preceding discussion,
may be characterized as a rational utilization of all possibilities of
unambiguous interpretation of measurements, compatible with the finite and
uncontrollable interaction between the objects and the measurements in the
field of quantum theory. \cite[p.700]{Bohr2}
\end{quotation}

\noindent There would be no difficulty in measuring the position of a
shutter as well as its momentum, so performing an inclusive measurement,
were it not for this local notion of an uncontrollable disturbance on
measurement. Quite distinct from this is the no-separation principle, the
non-local sense in which a phenomenon is defined relative to one or another
of such mutually exclusive experiments.

All this being so, why did Bohr have any difficulty with the EPR\ paper?
Rosenfeld reported that it was not all plain sailing, or not for the first
few days anyway \cite{Rosenfeld}. The argument surely required a different
answer to the one Bohr had presumably found to Einstein's earlier attempt to
draw out the non-local import of quantum mechanics, as also reported by
Rosenfeld\footnote{%
According to Rosenfeld \cite{Rosenfeld2}, Einstein posed an outline of the
EPR argument in Bruxelles in 1933, apparently only to \textquotedblleft
illustrate the unfamiliar features of quantum phenomena\textquotedblright .
(In fact the outline is of a rather different, and fallacious argument,
similar to the one that Dickson has criticized as an - unattrubuted - 
\textit{misreading }of the EPR argument \cite{Dickson1}.)}; or the answer he
had found to Einstein's \textquotedblleft photon box\textquotedblright\
thought experiment, another precursor to the EPR\ argument, at the 6th
Solvay conference. And the mathematical example given in the EPR\ paper,
making use of Dirac delta-function normalization and Fourier transforms of
two-particle wave functions, was hardly physically transparent. It made no
reference to dynamics, and it was not interpreted in terms of any actual or
possible experiment. Bohr undoubtedly had to struggle to find an
experimental model for it.

There is one last and crucial component to Bohr's reasoning. I have
mentioned a puzzle, following his insistence on the use only of classical
concepts in interpreting experiments. How after all does he get beyond
classical theory? How to express what cannot be expressed classically? For
Bohr, the quantum formalism itself has entirely disappeared from view.%
\footnote{%
Equations of quantum mechanics appeared only once in his reply to the EPR
argument, and that was in a footnote (presenting the two choices of
commutators, for relative positions or total momenta, in terms of the
transformation theory).} In what sense, then, was there any possibility of
genuinely non-classical laws? The answer, in a word, is \textit{%
complementarity}. Bohr immediately continued:

\begin{quotation}
\noindent In fact, it is only the mutual exclusion of any two experimental
procedures, permitting the unambiguous definition of complimentary physical
quantities, which provides room for new physical laws, the coexistence of
which might at first sight appear irreconcilable with the basic descriptions
of science. It is just this entirely new situation as regards the
description of physical phenomena that the notion of \textit{complementarity}
aims at characterizing. \cite[p.700, emphasis original]{Bohr2}
\end{quotation}

\noindent Bohr was scarcely under pressure to explain how this was to be
done (how to make use of this \textquotedblleft new room\textquotedblright
); he had showed it by example in the Como lecture and in the discussion
just concluded in reply to the EPR\ argument. It is the idea that this
amounted to a \textit{general }new method in the sciences that is the
crucial one.

This idea also preceded the EPR argument. It was clear from Bohr's address
to the Scandinavian Meeting of Natural Scientists in 1929, which first
appeared in German in 1931 and in English as \textquotedblleft The atomic
theory and the fundamental principles underlying the description of
nature\textquotedblright\ in 1934 \cite[pp.102-19]{Bohr1}. In this lecture
he proposed that complementarity might apply also to \textquotedblleft the
more profound biological problems\textquotedblright ; there
\textquotedblleft we must expect to find that the recognition of
relationships of wider scope will require that the same conditions be taken
into consideration which determine the limitations of the causal mode of
description in the case of atomic phenomena\textquotedblright . The point
was made more starkly in an Addendum to his \textquotedblleft Introductory
Survey\textquotedblright ,\ that he added in 1931: \textquotedblleft \textit{%
the strict application of those concepts which are adapted to our
description of inanimate nature might stand in a relationship of exclusion
to the consideration of the laws of the phenomena of life}%
\textquotedblright\ \cite[p.22-23, emphasis original]{Bohr1}. He suggested
that the same may apply to psychological laws (or rather the opposition of
psychological and physical laws), and that the recognition of this
\textquotedblleft will enable us to comprehend...that harmony which is
experienced as free will and analyzed in terms of
causality\textquotedblright . In his Faraday lecture of 1930 \cite{Bohr7},
Bohr suggested that the concepts of thermodynamics were mutually exclusive
of the concepts of statistical thermodynamics, and that complementarity
could be applied to that domain as well.

Complementarity, it was clear by the end of the 1920s, was for Bohr a novel
explanatory framework, more general than the traditional one of causal
space-time descriptions, that applied in principle to any empirical domain
in which concepts could only be applied under mutually exclusive
experimental conditions, providing room for the discovery and definition of
entirely new laws, despite their restriction to those self-same concepts.

\section{BOHR'S LATER PHILOSOPHY}

Bohr gave a lengthy commentary on the Como lecture and the subsequent
history of his discussions with Einstein in 1949, in his contribution to the
Schilpp volume of the \textit{Library of Living Philosophers }devoted to
Einstein. It is considered by many as the most authoritative of Bohr's
writings on quantum mechanics. In the argument for complementarity two
assumptions were now highlighted. For the first (the \textquotedblleft
necessity of classical concepts\textquotedblright ):

\begin{quotation}
\noindent \noindent \textit{However far the phenomena transcend the scope of
classical physical explanation, the account of all evidence must be
expressed in classical terms. }\noindent The argument is simply that by the
word \textquotedblleft experiment\textquotedblright\ we refer to a situation
where we can tell others what we have done and what we have learned and
that, therefore, the account of the experimental arrangement and of the
results of the observations must be expressed in unambiguous language with
suitable application of the terminology of classical physics.\cite[p.208,
emphasis original]{Bohr3}
\end{quotation}

\noindent For the second (the \textquotedblleft no-separation\
principle\textquotedblright ):

\begin{quotation}
\noindent \noindent \textit{The impossibility of any sharp separation
between the behaviour of atomic objects and the interaction with the
measuring instruments which serve to define the conditions under which the
phenomena appear }\cite[p.209, emphasis original]{Bohr3}
\end{quotation}

\noindent The argument for complementarity now depends on the
\textquotedblleft individuality\textquotedblright\ of quantum effects (but
Bohr had used just this term in his statement of the quantum postulate in
the Como lecture):

\begin{quotation}
\noindent In fact, the individuality of the typical quantum effects finds
its proper expression in the circumstance that any attempt of subdividing
the phenomena will demand a change in the experimental arrangement
introducing new possibilities of interaction between objects and measuring
instruments which in principle cannot be controlled. Consequently, evidence
obtained under different experimental conditions cannot be comprehended
within a single picture, but must be regarded as complementary in the sense
that only the totality of the phenomena exhausts the possible information
about the objects. \cite[p.209]{Bohr3}
\end{quotation}

\noindent Bohr shortly after continued, driving home the possibility of
novelty despite the restriction to classical concepts:

\begin{quotation}
\noindent While the combination of these concepts into a single picture of a
causal chain of events is the essence of classical mechanics, room for
regularities beyond the grasp of such a description is just afforded by the
circumstance that the study of the complementary phenomena demands mutually
exclusive experimental arrangements. \cite[p.210]{Bohr3}
\end{quotation}

\noindent The first and second assumptions were both stated explicitly (and
the second also italicized) in the \textquotedblleft Introductory
Survey\textquotedblright\ of 1929; the stated argument for complementarity
can be found in marginally different forms in every one of the publications
we have considered; the argument for the possibility of novelty was clearly
stated in 1929, and again in 1930 and 1931. These principles and these
arguments were the central ones in his reply to the EPR argument. Commenting
on the latter, Bohr again declared \textquotedblleft we are here dealing
with problems of just the same kind as those raised by Einstein in previous
discussions\textquotedblright\ \cite[p.231]{Bohr3}

The charge that Bohr's views underwent a radical change as a consequence of
the EPR argument, and in particular that the holism of object-apparatus were%
\textit{\ later} developments of Bohr's though (e.g. \cite[p.32-33]{Cushing}%
\cite[p.185]{Faye}), cannot be sustained. But that is not to say that his
theory did not have other failings. It made no direct reference to the
mathematical formalism of quantum mechanics, so there is plenty of ambiguity
in how to set it out as a formal interpretation.\footnote{%
It left open, in particular, the possibility that complementarity applied
equally to any pair of canonically conjugate variables (and that energy and
momentum \textit{vs} spacetime coordination is only one example among many).
\par
Bohr did of course acknowledge that from the transformation theory and
non-commutativity quite generally it followed \textquotedblleft that it is
never possible, in the description of the state of a mechanical system, to
attach definite values to both of two canonically conjugate
variables\textquotedblright\ \cite[p.696]{Bohr2}. He also acknowledged that
\textquotedblleft in the quantum mechanical description our freedom of
constructing and handling the experimental arrangement finds its proper
expression in the possibility of choosing the classically defined paramters
entering in any proper application of the formalism\textquotedblright\ \cite[%
p.229]{Bohr3}. It does not follow that energy-momentum \textit{vs }spacetime
coordination do not have a special significance in his theory. It was
exclusively these that were at issue in every argument and example of
complementarity that he gave in the texts that we have considered. (He made
two mentions of the transformation theory in the Como lecture, in neither
case linking it to complementary; one other in his reply to EPR,\ already
mentioned; and none other in his later writings.)} Classical descriptions,
within definite latitudes - as given by his quasi-classical formulation of
the uncertainty relations - replaced quantum ones, in a procedure of
doubtful generality; and it admitted, without comment or explanation, the
non-local sense in which a phenomenon is defined by its context. Finally,
the latter continued to be subsumed under the notion of \textquotedblleft
interaction\textquotedblright , without comment or qualification. This was
true even in his most careful statement of complementarity in 1958:

\begin{quotation}
\noindent Far from restricting our efforts to put questions to nature in the
form of experiments, the notion of complementarity simply characterizes the
answers we can receive by such inquiry, whenever the interaction between the
measuring instruments and the objects forms an integral part of the
phenomena. \cite[p.4]{Bohr4}
\end{quotation}

In fact all Bohr's attempts - half-hearted at best - to derive his
conclusions from independent and precisely stated hypotheses must be judged
failures. Witness, in 1949, the founding principle, \textquotedblleft the%
\textit{\ impossibility }of any sharp separation between the behaviour of
atomic objects and the interaction with the measuring instruments\textit{%
\textquotedblright }; and again in the very same publication, in summarizing
the lesson of a variety of thought experiments, that \textquotedblleft the
main point here is the \textit{distinction} between the objects under
investigation and \ the measuring instruments which serve to define, in
classical terms, the conditions under which the phenomena
appear.\textquotedblright\ Bohr had put the latter point even more strongly
in his reply to EPR:

\begin{quotation}
\noindent The necessity of discriminating in each experimental arrangement
between those parts of the physical system considered which are to be
treated as measuring instruments and those which constitute the objects
under investigation may indeed be said to form a principal distinction
between classical and quantum-mechanical description of physical phenomena. 
\cite[p.150]{Bohr2}
\end{quotation}

\noindent It is not at all clear, at this point, just what the assumptions
of his theory of complementarity really are: the impossibility of making a
sharp separation; the necessity for making a sharp separation; and somewhere
in this, the quantum postulate.

In Bohr's final paper on the subject, he first emphasized the foundation of
any kind of unambiguous physical evidence in the formation of permanent
marks \textquotedblleft such as a spot on a photographic plate caused by the
impact of an electron\textquotedblright , by processes of \textquotedblleft
irreversible amplification\textquotedblright . The necessity of a
object-apparatus divide now follows as a consequence of his earlier
stipulation of 1935:

\begin{quotation}
\noindent In all such points, the observation problem of quantum physics in
no way differs from the classical physical approach. The essentially new
feature in the analysis of quantum phenomena is, however, the introduction
of a \textit{fundamental distinction between the measuring apparatus and the
objects under investigation}. This is a direct consequence of the necessity
of accounting for the functions of the measuring instruments in purely
classical terms, excluding in principle any regard for the quantum of
action. \cite[p.3-4]{Bohr4}
\end{quotation}

\noindent In the same essay we read that the fundamental reason why quantum
indeterminism cannot be read as a species of classical statistical mechanics
is:

\begin{quotation}
\noindent In the case of quantum phenomena, the unlimited divisibility of
events implied in such an account is, in principle, excluded by the
requirement to specify the experimental conditions. Indeed, the feature of
wholeness typical of proper quantum phenomena finds its logical expression
in the circumstance that any attempt at a well-defined subdivision would
demand a change in the experimental arrangement incompatible with the
definition of the phenomena under investigation.\cite[p.4]{Bohr4}
\end{quotation}

\noindent Natural or not, there is no contradiction between these
principles. The latter insists there is no object without a context, the
former that the context is purely classical - and has to be specified as such%
\footnote{%
Or as rigidly connected to bodies of arbitrarily large mass.} to determine
an unambiguous spacetime coordination of the phenomenon. All experiments
(what Bohr also called the \textquotedblleft proper\textquotedblright\
measurement instruments \cite[p.221]{Bohr3}) must ultimately be described in
terms of their arrangement in space and time. As Bohr added:

\begin{quotation}
\noindent \lbrack T]he ascertaining of the presence of an atomic particle in
a limited space-time domain demands an experimental arrangement involving a
transfer of momentum and energy to bodies such as fixed scales and
synchronized clocks, which cannot be included in the description of their
functioning, if these bodies are to fulfil the role of defining the
reference frame. \cite[p.5]{Bohr4}
\end{quotation}

Bohr does I think have a consistent message, although there were pitfalls in
decoding it. He was never able to set it out deductively; he never succeeded
in defining the quantum postulate or his other attempted principles on a
clear phenomenological basis, on a par with the light postulate of special
relativity (the paradigm he clearly sought to emulate). But to my mind much
the most important of his failings is that he did not clearly acknowledge
that at most complementarity was a \textit{conjecture}; that there are new 
\textit{possibilities} for the definition of \textquotedblleft the
phenomena\textquotedblright\ opened up by complementarity that \textit{it
may or may not be possible }to integrate within a causal spacetime picture.
There is nothing in his arguments to show that that is impossible, or that
the quantum postulate must always remain an inscrutable (\textquotedblleft
irrational\textquotedblright ) foundation to the theory. Bohr had only a%
\textit{\ theory} that there could be no such theory.

\section{COMPLEMENTARITY AND THE FORMALISM OF QUANTUM MECHANICS}

I am in agreement with Scheibe, in his much-admired review of Bohr's
philosophy, that \textquotedblleft there is no\textit{\ single} formulation
of quantum mechanics based entirely and consistently on the principles
proposed by Bohr\textquotedblright , but I disagree when he goes on to say:

\begin{quotation}
\noindent \lbrack W]e have in fact no technically elaborated formulation of
Bohr's quantum mechanics and that \textit{consequently} it is not at present
possible to make a really useful assessment of his
contributions.\textquotedblright \cite[p.5, emphasis original]{Scheibe}
\end{quotation}

\noindent Bohr did not so much as \textit{attempt} to give a formulation of
quantum mechanics, of the sort that Scheibe was interested in; now was he
interested in explaining the formalism. He was interested in the qualitative
phenomena.

Bohr did not write down the Schr\"{o}dinger equation in any of the writings
we have considered; the canonical commutation relations only twice. The
formalism, only once mentioned in his extensive review of 1949, was
\textquotedblleft an adequate tool for a complementary way of
description\textquotedblright , it \textquotedblleft represents a purely \
symbolic scheme permitting only predictions\textquotedblright . In 1958,
that in the Schr\"{o}dinger equation \textquotedblleft we are here dealing
with a purely symbolic procedure, the unambiguous physical interpretation of
which in the last resort requires a reference to a complete experimental
arrangement\textquotedblright . The \ formalism has a \textquotedblleft
non-pictorial character\textquotedblright . The fact that the wave-function
is defined on configuration space rather than position space, and the
essential role played by complex numbers in it, were cited by Bohr (from the
Como lecture on) as reasons to view it as \textit{purely} abstract (a point
of view shared by de Broglie).

Present at the Warsaw conference of 1938 was John von Neumann. Like Bohr, he
gave an address on the interpretation of quantum mechanics, but very
different in spirit. He gave an axiomatization of quantum mechanics as a
projective geometry, over one of the reals, complex numbers or quaternions,
in terms of purely lattice-theoretic axioms. They were in turn interpreted
as the expression of new logical laws. Indeterminacy, \ in the quantum
mechanical sense, was related (so he argued) to the failure of
distributivity. Von Neumann promised further connections with the continuous
geometries that had arisen in his recent work on the classification of
operator algebras.

For Bohr it was a stretch. But we learn something of his attitude to
complementarity from his reply, as reported thus:

\begin{quotation}
\noindent We must also notice that the question of the logical forms which
are best adapted to quantum theory is in fact a practical problem, concerned
with the choice of the most convenient manner in which to express the new
situation that arises in this domain. Personally, he compelled himself to
keep the logical forms of daily life to which actual experiments were
necessarily confined. The aim of the idea of complementarity was to allow of
keeping the usual logical forms while procuring the extension necessary for
including the new situation relative to the problem of observation in atomic
physics. \cite[p.xxx]{Bohr6}
\end{quotation}

\noindent Bohr was prepared to acknowledge the possibility of a revised
logic, and alternatives to complementarity. But as usual he was dogmatic
about the impossibility of a classically visualizable interpretation of
quantum phenomena:

\begin{quotation}
\noindent \lbrack I]t seems likely that the introduction of still further
abstractions into the formalism will be required to account \ for the novel
features revealed by the exploration of atomic processes of very high
energy. The decisive point, however, is that in this connection there is no
question of reverting to a mode of description which fulfils to a higher
degree the accustomed demands regarding pictorial representation of the
relationship between cause and effect. \cite[p.xxx]{Bohr6}
\end{quotation}

His case was unproven, however. The pilot-wave theory is a counterexample.
Why could there be no question of reverting to determinism? It is a disaster
for our understanding of Bohr and his doctrines that Bohr never made any
public statement on it.

On the measurement problem Bohr was almost as evasive. Bohr never chose to
address it as it arises at the formal level, in terms of the contrast
between the projection postulate and the unitary equations of motion. There
is only one comment he made in print that we can be sure was directed at the
problem in this sense. At the same Warsaw conference, Bialobizeski, the
President of the University and the chair of the conference, reviewed,
somewhat imperfectly, von Neumann's formulation of the measurement problem
along these formal lines. He suggested that the duality between the unitary
equations and the projection postulate could \ not be assimilated to the
choice of one or another of two complementary descriptions, and that it must
remain a fundamental posit of the theory. Bialobizeski surely had a point.
Here is the report of Bohr's response:

\begin{quotation}
\noindent \lbrack T]he duality he noticed in the interpretation of the
formalism of quantum \ mechanics was, in his opinion, a question of choosing
the most adequate description of the experiment. If we decide to include in
the enumeration of the exterior conditions all the instruments which must be
used for the study of the whole phenomenon, the only arbitrary factor
remaining is, as he had explained in his paper, the free choice of these
experimental conditions, and, apart from this freedom, the interpretation of
the solution of the problem, concerning the predictions to which the
phenomenon we are studying leads, is perfectly unequivocal.\cite[p.xxx]%
{Bohr6}
\end{quotation}

\noindent The comment is too cryptic to be really illuminating. One can read
a plausible story into it: it is that the free choice of the experimental
conditions determines, as a choice of the parts of the apparatus in space
and time (the positions of diaphragms, mirrors, springs and what have you),
the relevant basis (the classical variables defined with what latitudes),
and that once this is fixed, the probabilistic interpretation is unequivocal
- just as effected by the projection postulate (as a device to define
probabilities from the wave-function). Wave-packet collapse, real (as in a
stochastic theory) or effective (as in pilot-wave theory or in the Everett
interpretation) can then be identified with the individual and holistic
element of the quantum postulate as well as with its uncontrollable element
- but this is all to put the matter in terms Bohr could hardly have
recognized. If Bialobizeski had pressed the suggestion that the unitary
equations also explicitly incorporate the choice of experiment, the deeper
objection that Bohr was really committed to was that\textit{\ the unitary
dynamics means nothing at all in itself}. That there \textit{is} no causal,
spacetime description thus provided that has to be suspended, when a
measurement is performed. The quantum formalism is \textit{only} an abstract
calculus. As we have seen, Bohr made this point over and over again.

\section{COMPLEMENTARITY AS A CONJECTURE}

Complementarity, as I read it, was a conjecture of unusually wide scope:
that where the experimental definition of certain concepts precluded the
definition of others, new regularities could be defined, that could not be
captured by any unified treatment involving all of them. Similarly, one
might conjecture that cosmology as an empirical discipline is impossible; or
that science as a strictly unified discipline is impossible. Closer to home
is Rovelli's recent conjecture \cite{Rovelli}, that whilst any system can be
modelled piecemeal in quantum mechanics, no model can be given of the
totality of physical systems; or conjectures by Fuchs and others \cite{Fuchs}%
\cite{Zeilinger} on the possibilities of encoding information and the
transfer of information. Bohr has plenty of emulators today.

I see no \textit{a priori} principle that rules out any of these strategies,
but that is only to say that they are at the end \textit{scientific }%
conjectures, albeit at an unusually high level of abstraction; they stand or
fall by their durability and success. When it comes to complementarity,
surely, the jury is by now finally in.

In its negative claims complementarity denied the possibility of a causal
spacetime explanation for key experiments of quantum mechanics. In addition,
Bohr insisted that the formalism can only be interpreted by specification of
a (classically defined) context of measurement. But there are now plenty of
examples of causal spacetime explanations for the phenomena that Bohr
considered (as given in all the major realist schools today, whether
pilot-wave theory, GRW theory, or the Everett interpretation); and we have
in decoherence theory techniques for obtaining approximately classical
descriptions from quantum ones that evade Bohr's strictures entirely.

On the positive side, Bohr did offer a framework for the analysis of quantum
phenomena in terms of classical concepts. Here we may grant him limited
success. The key point is that he offered an avenue - a new method in
science - whereby apparently inconsistent explanations could be reconciled.
But it would have been \textit{ad hoc} to restrict it to quantum physics; if
a new scientific method, it should have applications in other fields as
well. On two counts progress was expected: progress with complementary
theorizing in quantum physics - reasoning along the lines he laid out in the
Como lecture, in his reply to EPR, and in his 1949 review - and progress
with a similar style of theorizing in other disciplines, whose elementary
concepts also stand in such a relation of mutual exclusion. Neither has been
forthcoming.\footnote{%
Bohr, in collaboration with Rosenfeld in 1933 did gave one other example of
the genre (an operational analysis of the limits to definability in free
quantum electromagnetic field theory \cite{Bohr8}); and later, in 1950, an
attempt at a comparable study of the interacting theory \cite{Bohr11}. But
the panalophy of levers, trapdoors and springs so introduced seemed little
sort of baroque.
\par
Bohr was surely disappointed by the failure of complementarity in other
fields. Whilst he still held out the hope of applications of complementarity
to biology and the social sciences in 1958, he was markedly less optimistic
in his final comments on the subject in 1962, a few months before his death.
There he acknowledged that the use of teleological explanations in biology
did not in fact imply any restriction on the application of physics to that
field, adding, in a departure from his text, that \textquotedblleft in the
last resort, it is a matter of how one makes headway in biology. I think
that the feeling of wonder which the physicists had thirty years ago has
taken a new turn. Life will always be a wonder, but what changes is the
balance between the feeling of wonder and the courage to try to
understand\textquotedblright\ \cite[p.26]{Bohr12} (The implication, that
with complementarity one did not have the courage to understand, was surely
unintended.)}

The historical record is clearly negative. But these failures could not have
been foreseen in the 1920s and '30s. In retrospect, it is clear that Bohr's
philosophy was essentially conservative, an extension of the methods that
had served him and his contemporaries well in the creation of quantum
mechanics. If a competing approach to the theory required their abandonment,
without benefit in terms of new experiments, little wonder that Bohr won the
day: progress by increment, by conservative extension of existing concepts
and fragments of theories, may not exactly be a principle of rationality,
but \textit{pace} Popper and Kuhn, it is has served dynamics well through
its long history \cite{Saunders2}.

I come back to the pilot-wave theory. At certain points I have said that it
is a counter-example to Bohr's claims. How then could he have ignored it?
Bohr's negligence in this respect may seem remarkable - so much so that it
puts in question our whole reading of his philosophy. But it is much worse
on Heisenberg's reading, that the \textquotedblleft
Copenhagen\textquotedblright\ interpretation, as he called it, was based on
the need to introduce \textquotedblleft the observer\textquotedblright\ in
physics;\footnote{%
Heisenberg's reading of orthodoxy made a happy target, in different ways,
for Popper, Feyerabend, and Hanson, who thereby helped to popularize it (I
am indebted to Don Howard for this and other observations on this history.)}
or on von Neumann's, that the measurement postulates only reflected the fact
that any account of the objective world must ultimately terminate in \textit{%
conscious }perception. Claims of this scope are \textit{straightforwardly}
refuted by the pilot-wave theory.

Consider again the conservatism of Bohr's assumptions. He insisted on no
philosophical principle, unless it was a broad and shallow operationalism.
He was dogmatic only in regard to the indispensability of classical
concepts. He offered cogent reasons to believe that the classical ideal of
explanation, as a causal spacetime description, may not be available in the
quantum domain. In its place he provided simple, context-relative accounts
of paradigmatic quantum phenomena in terms of classical concepts, many of
them simple extensions of the ideas of the old quantum theory. Against it
pilot-wave theory made of these quantities (apart from position), in any
dynamical context, nothing but \textit{epiphenomena}, artifacts of the
kinematical limit. Focus instead on the \textit{real} motions in
configuration space, and all the usual interactions of electrodynamics
appear to be something close to \textit{nonsense}.\footnote{%
For example, the electron may be at rest within the atom, despite the
Coulomb interaction; a photon reflected from a mirror may be perfectly at
rest. (These examples were pointed out by Brillouin at the 5th Solvay
conference \cite[p.120, 137-40, 266]{Solvay}.)} It made nonsense of some of
the new equations as well \ (for example, the uncertainty relations):
considered as \textquotedblleft a principle which is necessary for the
interpretation of the\textit{\ totality} of quantum
phenomena\textquotedblright

\begin{quotation}
\noindent These results, which the coherence and experimental verification
of the new mechanics have placed beyond any doubt, can in no way be
reconciled with the pilot-wave theory. The latter leads in particular to a
well-defined value of the linear momentum and does not allow us to obtain
the uncertainty relations. \cite[p.177]{de Broglie}
\end{quotation}

\noindent The remark was made by none other than de Broglie, a quarter of a
century \textit{after} the 5th Solvay conference; he allowed it to stand
even \textit{after }reading Bohm's papers of 1952. It shows how hard it was
for him to accept that the dynamical quantities revealed by experiment were
not the real ones outside of the measurement context, and \textit{vice versa}%
.

Total energy, kinetic energy, momentum and spin as revealed by measurement
are not in general to be found in the particle trajectories \cite{Brown1}.
The pilot-wave theory is contextual with respect to all of them, in the
technical sense, as required by the Kochen-Specker-Bell theorem, but (with
the possible exception of spin) they are contextualized in Bohr's sense too
- which of such variables has what meaning (what \textquotedblleft
latitude\textquotedblright ) is a matter of the experimental context. And
what the pilot-wave theory does \textit{not} deliver is a causal, spacetime
account \textit{of those very variables, across all contexts} - just because
they do not attach to the particle trajectories. In this sense, it may even
be said, the pilot-wave theory is no counter-example to the principle of
complementarity.

\bigskip

$\mathbf{ACKNOWLEDGEMENTS}$

\bigskip

My thanks to Don Howard and Michael Dickson for illuminating discussions of
Bohr's interpretation (although they cannot be held responsible for the
particular light I have tried to shed on it), and to Guido Bacciagaluppi
and, especially, to Antony Valentini, for any number of historical
corrections (they are not at fault for any remaining ones!). Above all my
thanks to the editors for the opportunity to express my thanks and
appreciation for the life and work of Jim Cushing; for his friendship and
support in difficult times, and for his fearless questioning in the easy
ones.

\end{document}